\documentstyle[aps]{revtex}
\begin{document}
\draft
\title{
Comment on 
$\lq\lq$Theory of Hall Effect and Electrical Transport 
in High-$T_{\rm c}$ Cuprates: 
Effects of Antiferromagnetic Spin Fluctuations"
\cite{N}
}
\author{
O. Narikiyo
}
\address{
Department of Physics, 
Kyushu University, 
Fukuoka 810-8560, 
Japan
}
\date{
December, 2000
}
\maketitle
\begin{abstract}
I point out 
that a magnetotransport theory 
on the basis of the fluctuation exchange (FLEX) approximation, 
which is proposed in order to explain anomalous transport properties 
in the normal state of high-$T_{\rm c}$ cuprate superconductors, 
is not compatible with the Fermi liquid theory 
and that the anomaly obtained there 
ascribes to the inconsistency of the approximation. 
\end{abstract}
\vskip 18pt
  Recently a magnetotransport theory\cite{KK1} 
on the basis of the fluctuation exchange (FLEX) approximation 
is proposed in order to explain anomalous transport properties 
in the normal state of high-$T_{\rm c}$ cuprate superconductors. 

  In this comment I point out 
that such an approximation cannot be allowed 
from the view point of the Fermi liquid theory in the following. 
  The same criticism is relevant to the series of papers.
\cite{KK2} 

\unitlength=1mm
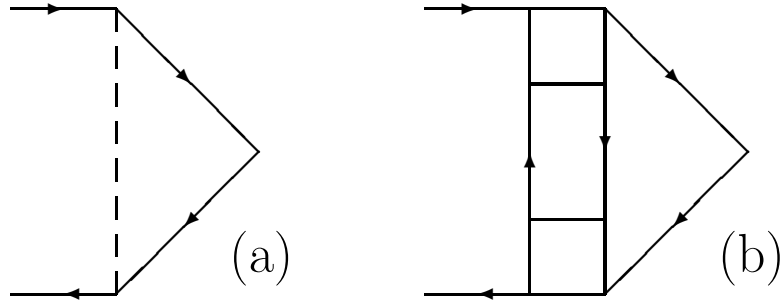
\begin{figure}
\begin{center}
\begin{picture}(140,50)
\thicklines
\put(50,5){\huge(a)}
\put(21,40){\vector(1,0){7}}
\put(28,40){\line(1,0){7}}
\put(35,40){\vector(1,-1){10}}
\put(45,30){\line(1,-1){9}}
\put(54,21){\vector(-1,-1){10}}
\put(44,11){\line(-1,-1){9}}
\put(35,2){\vector(-1,0){7}}
\put(28,2){\line(-1,0){7}}
\multiput(35,40)(0,-5){8}{\line(0,-1){3}}
\put(115,5){\huge(b)}
\put(76,40){\vector(1,0){7}}
\put(83,40){\line(1,0){7}}
\put(90,40){\line(1,0){10}}
\put(100,40){\vector(1,-1){10}}
\put(110,30){\line(1,-1){9}}
\put(119,21){\vector(-1,-1){10}}
\put(109,11){\line(-1,-1){9}}
\put(100,2){\line(-1,0){10}}
\put(90,2){\vector(-1,0){7}}
\put(83,2){\line(-1,0){7}}
\put(90,30){\line(1,0){10}}
\put(90,30){\line(0,1){10}}
\put(100,30){\line(0,1){10}}
\put(90,12){\line(1,0){10}}
\put(90,2){\line(0,1){10}}
\put(100,2){\line(0,1){10}}
\put(90,12){\vector(0,1){9}}
\put(90,21){\line(0,1){9}}
\put(100,30){\vector(0,-1){9}}
\put(100,21){\line(0,-1){9}}
\end{picture}
\caption{
The lowest order current vertex correction 
(a) in the FLEX approximation and 
(b) in the Fermi liquid theory.
}
\end{center}
\end{figure}

  In the FLEX magnetotransport theory\cite{KK1} 
the current vertex is constructed as Fig.\ 1(a) 
where only the lowest order process is shown. 
  Here the broken line represents the susceptibility 
of the spin fluctuation. 
  Higher order processes are constructed by the ladder 
series of this process. 
  If the antiferromagnetic correlation is enhanced, 
this series gives a repeated scattering between 
the state with wavenumber ${\bf k}$ and ${\bf k+Q}$ 
where ${\bf Q}=(\pi/a,\pi/a)$ for the square lattice 
of the lattice constant $a$. 
  This repeated scattering with large wavenumber change 
is the origin of an anomalous transport properties 
in the FLEX approximation.\cite{KK1,KK2}

  On the other hand, in the Fermi liquid theory 
the current vertex should be constructed as Fig.\ 1(b) 
where the empty square represents the 4-fermion interaction 
vertex $\Gamma$. 
  It should be noted that the fermion propagator 
represented by the line with the arrow corresponds to 
the coherent part of the propagator. 
  Here the propagator is related to the spectral function 
$\rho_{\bf k}(\omega)$ and this function is decomposed as 
\begin{equation}
\rho_{\bf k}(\omega)
=\rho_{\bf k}^{\rm coh}(\omega)+\rho_{\bf k}^{\rm inc}(\omega).
\end{equation}
  The coherent part $\rho_{\bf k}^{\rm coh}(\omega)$ is 
the spectral weight of the asymptotic free fermion and 
the incoherent part $\rho_{\bf k}^{\rm inc}(\omega)$ is 
that of the scattered state. 
  In the Fermi liquid theory 
the contribution of the incoherent part is confined 
in the interaction vertex $\Gamma$ and this interaction 
is defined only for the coherent part of the fermion. 
  This structure is dropped in the FLEX approximation. 
  If the above repeated scattering is dominant, 
the spectral function $\rho_{\bf k}(\omega)$ is dominated 
by the incoherent part $\rho_{\bf k}^{\rm inc}(\omega)$. 
  In this case the renormalized amplitute $z$ of the asymptotic 
free fermion is vanishingly small. 
(This vanishing $z$ might be related to the so-called pseudogap.) 
  Even if the interaction vertex $\Gamma$ is enhanced, 
the effective interaction for the quasiparticle $z^2\Gamma$ 
is not enhanced. 
  This feedback ensure the applicability of the Fermi liquid theory. 
  The FLEX approximation lacks this feedback and is not compatible 
with the Fermi liquid theory. 
  Specifically the FLEX approximation has no counterpart 
to the constraint that the vertical lines in Fig.\ 1(b) 
should be coherent. 
  So that Fig.\ 1(a) is an overestimation of Fig.\ 1(b). 

  Since the FLEX theory denies the established transport theory,\cite{YY} 
which is based on the Fermi liquid theory and consistent with 
the usual Boltzmann transport theory, 
we should be careful about FLEX's validity. 
  In the Fermi liquid theory\cite{YY} 
the effect of the current vertex correction cancels with the self-energy 
correction and leads to the conservation of the pseudomomentum. 
  The conservation law is independent of the interaction strength. 
  Moreover, it leads to the well-recognized masking factor, $1-\cos \theta$, 
in the case of elastic scattering. 
  These things are the components of the usual and established 
transport theory not found in the FLEX approximation. 

  In summary, 
I point out that the result of the FLEX approximation 
violates the rigorous relation which should be satisfied 
in the Fermi liquid. 
  The transport anomaly obtained in the FLEX approximation 
ascribes to the inconsistency of the approximation. 


\end{document}